# Applicability of the Rytov full effective-medium formalism to the physical description and design of resonant metasurfaces

HAFEZ HEMMATI[1,2] AND ROBERT MAGNUSSON[1,] [*]

[1]*Department of Electrical Engineering, University of Texas at Arlington, Arlington, Texas 76019, USA*
[2]*Department of Materials Science and Engineering, University of Texas at Arlington, Texas 76019, USA*
*\* magnusson@uta.edu*

**Abstract:** Periodic photonic lattices constitute a fundamental pillar of physics supporting a plethora of scientific concepts and applications. The advent of metamaterials and metastructures is grounded in deep understanding of their properties. Based on Rytov's original 1956 formulation, it is well known that a photonic lattice with deep subwavelength periodicity can be approximated with a homogeneous space having an effective refractive index. Whereas the attendant effective-medium theory (EMT) commonly used in the literature is based on the zeroth root, Rytov's closed-form transcendental equations possess, in principle, an infinite number of roots. Thus far, these higher-order solutions have been totally ignored; even Rytov himself discarded them and proceeded to approximate solutions for the deep-subwavelength regime. In spite of the fact that Rytov's EMT models an infinite half-space lattice, it is highly relevant to modeling practical thin-film periodic structures with finite thickness as we show. Therefore, here, we establish a theoretical framework to systematically describe subwavelength resonance behavior and to predict the optical response of resonant photonic lattices using the full Rytov solutions. Expeditious results are obtained because of the semi-analytical formulation with direct, new physical insights available for resonant lattice properties. We show that the full Rytov formulation implicitly contains refractive-index solutions pertaining directly to evanescent waves that drive the laterally-propagating Bloch modes foundational to resonant lattice properties. In fact, the resonant reradiated Bloch modes experience wavelength-dependent refractive indices that are solutions of Rytov's expressions. This insight is useful in modeling guided-mode resonant devices including wideband reflectors, bandpass filters, and polarizers. For example, the Rytov indices define directly the bandwidth of the resonant reflector and the extent of the bandpass filter sidebands as verified with rigorous simulations. As an additional result, we define a clear transition point between the resonance subwavelength region and the deep-subwavelength region with an analytic formula provided in a special case.

## 1. Introduction

Periodic photonic lattices, known as diffraction gratings for 100 years and diffractive optical elements for decades, have a venerable history [1-5]. With major discoveries in optical physics deriving from their deployment, periodic structures enable wide application fields including spectroscopy, laser technology, and sensors. Imbuing the lattice with waveguiding capability offers yet another set of functionalities grounded in resonance effects due to excitation of lateral leaky Bloch modes [6-14]. In the recent past, periodic photonic lattices are often referred to as "metasurfaces" or "metamaterials" in which periodically aligned wavelength-scale features enable manipulation of an incoming electromagnetic waves in a desired manner [15-19]. Resonant lattices offer novel properties and light-wave control in compact format potentially replacing and complementing conventional optical devices.

Extensive theoretical and experimental studies have been conducted to realize resonant and nonresonant periodic structures in materials systems pertinent to the various spectral regions. Whereas various wavelength ($\lambda$) to periodicity ($\Lambda$) ratios can be deployed, working in the subwavelength regime offers a particularly efficient optical response. Transition from the non-subwavelength to the subwavelength regime occurs at the Rayleigh wavelength ($\lambda_R$) [2]. For wavelength values longer than $\lambda_R$, all higher diffraction orders are eliminated and only the zero orders propagate in the cover and substrate media. In the subwavelength regime, one can define two main regions. These are the deep-subwavelength region where the wavelength is much larger than the period, showing thin-film effects on account of a high degree of homogenization, and the resonant subwavelength region where the wavelength-scale periodicity triggers guided-mode, or leaky-mode, resonance effects. These regions are shown schematically in Fig. 1(a). While the Rayleigh wavelength is known by $\lambda_R = n_S \Lambda$, there exists presently no definition for $\lambda_c$; we propose a definition for this value in this paper.

Since the seminal work by Rytov in 1956, the effective refractive indices of subwavelength gratings can be calculated for both transverse electric (TE) and transverse magnetic (TM) polarization states [20]. His effective-medium theory (EMT) applies to an infinite periodic halfspace. Treating continuity and periodicity of the electromagnetic fields at boundaries between constituent materials in a unit cell results in polarization-dependent transcendental equations. Employing a series expansion for the tangent term in the transcendental equations returns the well-known zero-order, second-order, or higher-order approximate solutions of effective refractive indices. Applying EMT based on the approximated Rytov formulation, one can replace a subwavelength grating by an equivalent homogeneous film with corresponding effective refractive indices for each polarization. This process is noted schematically in Figs. 1(b) and 1(c). The thickness of the homogeneous film is identical to the grating thickness. In the deep subwavelength, or quasi-static, limit $\lambda/\Lambda \to \infty$, the zero-order effective refractive indices result in a reliable solution in terms of equivalent reflection, transmission, and phase calculations. Notably, in 1986, Gaylord et al. implemented zero-order EMT to approximate subwavelength gratings with a single homogeneous layer in order to design an antireflection coating at normal incidence [21]. In a related work, Ono et al. approximated a sinusoidal ultrahigh spatial frequency grating by several rectangular grating layers with different fill factors to design an antireflection structure [22]. They calculated the refractive index of each rectangular layer using the zero-order approximation. However, the zero-order approximation fails for wavelengths outside the deep subwavelength regime. Therefore, as the value of $\lambda/\Lambda$ approaches the resonant subwavelength regime (i.e. $\lambda \sim \Lambda$) higher-order approximations must be used. Thus, Richter et al. used second-order EMT to design and study optical elements with a form birefringent structure [23]. Moreover, Raguin and Morris utilized second-order EMT to design antireflection surfaces in the infrared (IR) electromagnetic bands [24].

All previous EMT studies [21-29] based on Rytov's formulation [20], with either exact or approximated solutions, have reported only one effective refractive index for each wavelength as depicted in Fig. 1(d). In contrast, here, we report that solving the exact transcendental equation in the resonant subwavelength regime can result in several effective refractive indices for a single wavelength. Mathematically, since $\tan(x)$ has an infinite sets of roots, this may not come as a surprise. What is surprising is that these roots are highly applicable to practical problems modeling photonic lattices with finite thickness, namely metasurfaces and metamaterials, as we show in detail in the remainder of the paper. These higher-order solutions have been completely ignored thus far to our knowledge. Even Rytov himself paid no attention to them and proceeded to derive simplified approximate expressions based on the zeroth root [20]. In his case, this is understandable as resonant photonic lattices were not known at that time.

Henceforth, we establish our theoretical framework to systematically describe subwavelength resonance behavior and predict the optical response of resonant photonic lattices using the full Rytov solutions. Expeditious results are obtained because of Rytov's semi-analytical formulation with direct, new physical insights available for resonant lattice properties. To prove the correctness of the proposed approach, we compare our semianalytical results with rigorously computed results and show excellent agreement between them. Our solutions, based on the exact symmetric Rytov problem, are previewed schematically in Fig.

1(e). Most importantly, we show here that the higher Rytov solutions $n_m^{EMT}$ correspond exactly to reradiated fields generated by higher-order evanescent diffracted waves represented as $S_m$, m=±1, ±2, …, driving the resonance process [30-32].

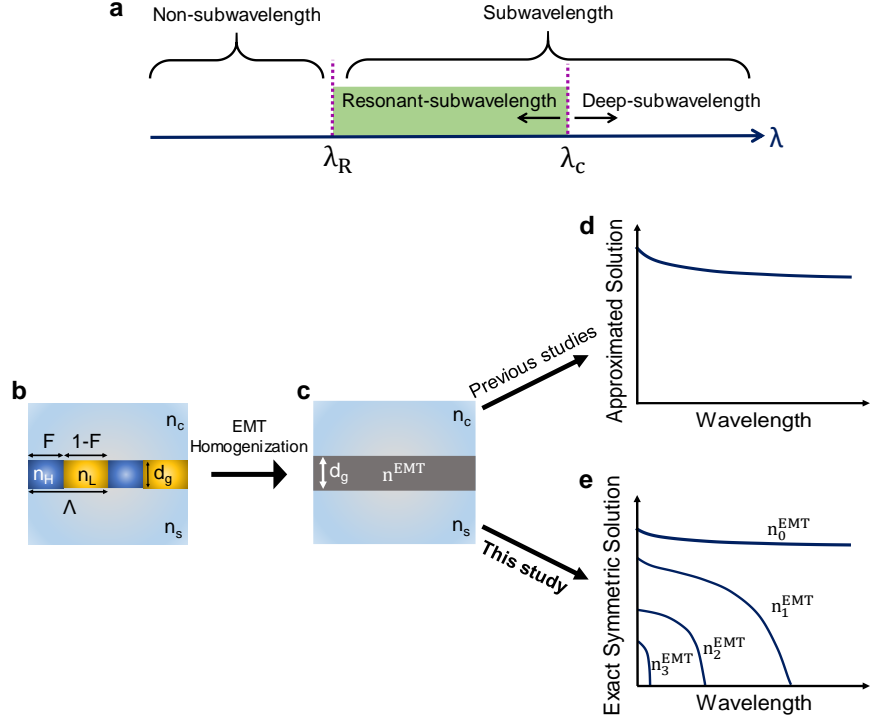

Fig. 1. Schematics illustrating (a) the diffraction regimes pertaining to Rytov's solutions, (b) the general rectangular grating model, (c) equivalent thin-film EMT model, (d) the zeroth-root Rytov solution basic to all past EMT models, (e) calculated effective refractive indices presented in this study based on the exact Rytov formalism.

## 2. Rytov refractive indices and their interpretation

We first review the Rytov formalism [20] for TE polarization, where the electric-field vector is parallel to the grating lines. The full formula for a rectangular grating structure with infinite thickness is derived by considering the continuity of the electric and magnetic fields at boundaries between the ridges and grooves. There results a transcendental equation given by

$$(1+\kappa^2)\sin(\alpha_1 a)\sin(\alpha_2 b) + 2\kappa(1-\cos(\alpha_1 a)\cos(\alpha_2 b)) = 0 \quad (1)$$

where $\alpha_1 = k_0\sqrt{n_H^2 - n^2}$, $\alpha_2 = k_0\sqrt{n_L^2 - n^2}$, $k_0 = 2\pi/\lambda_0$, and $\kappa = \alpha_1/\alpha_2$. The parameters a and b are the widths of the grating constituents with refractive indices of $n_H$ and $n_L$, respectively. Based on this, one can define parameters F=a/Λ and 1-F=b/Λ as a fill factors of each section in a unit cell as shown in Fig. 2(a). Since symmetric rectangular gratings are considered in Rytov's model, he extracted solutions from the full formula Eq. (1) that are pertinent to symmetric field distributions inside the grating. Accordingly, Eq. (1) is reduced to Eq. (2) which we reference here as the "exact" Rytov formulation for TE polarization.

$$\sqrt{n_L^2 - (n_{TE}^{EMT})^2}\tan\left[\frac{\pi\Lambda}{\lambda}(1-F)\sqrt{n_L^2 - (n_{TE}^{EMT})^2}\right] = -\sqrt{n_H^2 - (n_{TE}^{EMT})^2}\tan\left[\frac{\pi\Lambda}{\lambda}F\sqrt{n_H^2 - (n_{TE}^{EMT})^2}\right] \quad (2)$$

Similarly, for TM polarization, where the magnetic-field vector is parallel to the grating lines, there results

$$\frac{\sqrt{n_L^2 - (n_{TM}^{EMT})^2}}{n_L^2} \tan\left[\frac{\pi\Lambda}{\lambda}(1-F)\sqrt{n_L^2 - (n_{TM}^{EMT})^2}\right] = -\frac{\sqrt{n_H^2 - (n_{TM}^{EMT})^2}}{n_H^2} \tan\left[\frac{\pi\Lambda}{\lambda} F \sqrt{n_H^2 - (n_{TM}^{EMT})^2}\right] \quad (3)$$

Solving the exact Rytov equations, Eq. (2) and Eq. (3), for $n_{TE}^{EMT}$ and $n_{TM}^{EMT}$ delivers a set of effective refractive indices that depend on the wavelength and the input design parameters. In principle, due to the periodicity of tan(x) there exists an infinite number of solutions; in practice, a few of the lowest-order solutions will be useful. Except for $n_0^{EMT}$, the effective refractive indices have specific cutoff wavelengths. Knowing the cutoff wavelengths is key to predicting the optical response as shown here. Working at wavelengths longer than the Rayleigh wavelength $\lambda_R = n_S\Lambda$, ensures zero-order propagation towards the cover and substrate with all higher-order diffracted waves being evanescent. These higher diffraction orders propagate in the periodic region depending on the structural design and corresponding cutoff values of $\lambda_c^m$.

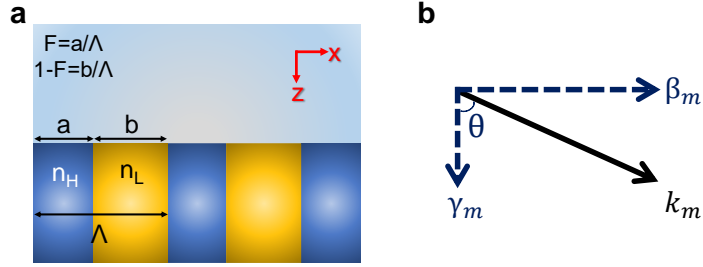

Fig. 2. (a) Schematic of the half space grating model in Rytov's formulation. (b) Wavevector of the m[th] diffracted order accompanied by its vertical and horizontal components.

In the periodic region, the fundamental coupled wave expansion of the y-component of the electric field can be written as [33, 34]

$$E_y(x,z) = \sum_m S_m(z) \exp[-i(k - mK)x] \quad (4)$$

where $S_m(z)$ are the amplitudes of the space-harmonic components in the Fourier series expansion of the total field in periodic direction, k is the wave vector of the diffracted wave, and $K = 2\pi/\Lambda$ is the grating vector magnitude. Each diffracted order possesses a wavevector ($k_m$) in the direction of propagation which can be resolved into vertical and horizontal components as depicted in Fig. 2(b). Effective refractive indices obtained by solving the Rytov equations are pertinent to the vertical components of the diffracted orders. We have

$$k_m^2 = \beta_m^2 + \gamma_m^2 \quad (5)$$

where $\beta_m = k_m \sin\theta$, $\gamma_m = k_m \cos\theta$, $k_m = k_0 n_m(\lambda_0)$, and $k_0 = 2\pi/\lambda_0$. Defining $N_m(\lambda_0) = \beta_m/k_0 = n_m(\lambda_0)\sin\theta$ and $n_m^{EMT}(\lambda_0) = \gamma_m/k_0 = n_m(\lambda_0)\cos\theta$, a relation is obtained between the component refractive indices of Fig. 2(b) as

$$\left(n_m(\lambda_0)\right)^2 = \left(N_m(\lambda_0)\right)^2 + \left(n_m^{EMT}(\lambda_0)\right)^2 \quad (6)$$

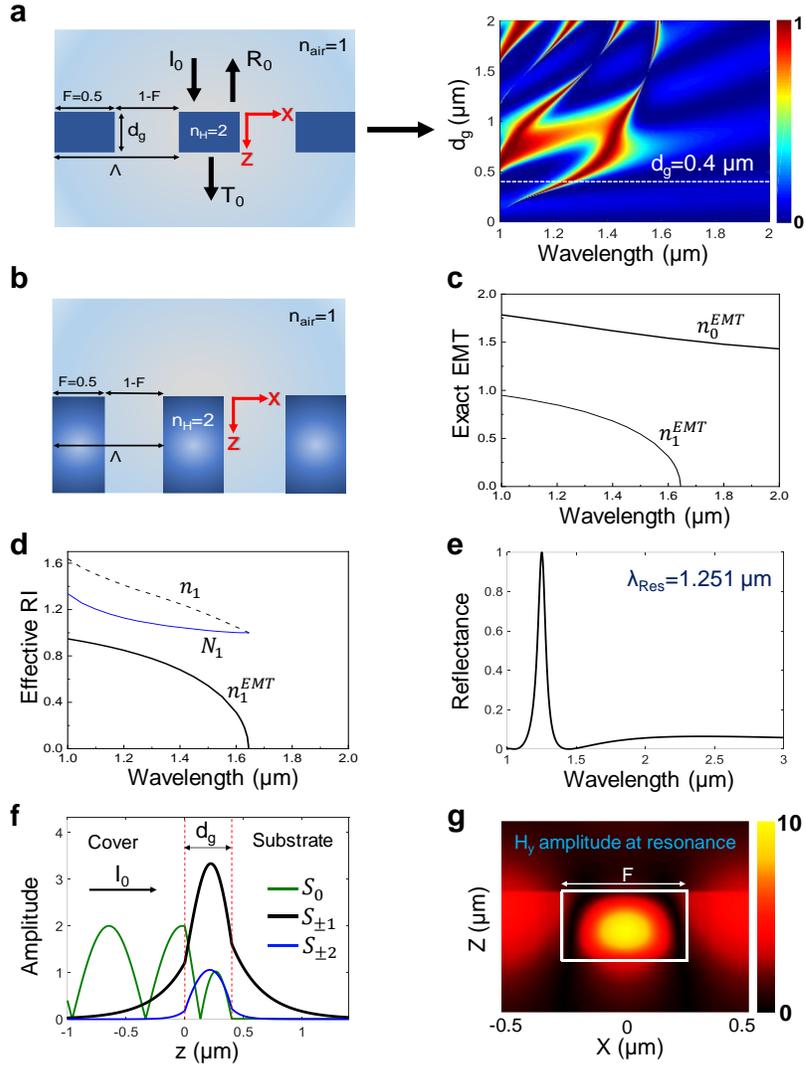

Fig. 3. (a) Schematic of a representative grating membrane and corresponding reflection map as a function of grating thickness ($d_g$) for TM-polarized incident light, (b) schematic of the half-space grating model, (c) calculated exact effective refractive indices, (d) wavelength dependent effective refractive indices of waveguide ($n_1$), horizontal component ($N_1$), and vertical component ($n_1^{EMT}$) based on Eq. (6), (e) simulated reflectance spectrum of a grating with $d_g =0.4$ μm, (f) amplitude of the coupled diffracted orders at resonance wavelength of $\lambda_{Res}=1.251$ μm, and (g) distribution of total magnetic field in one period at the resonance wavelength of $\lambda_{Res}=1.251$ μm showing $TM_0$ mode shape. The grating structure has constant parameters of $\Lambda=$ 1 μm, F=0.5, $n_H$=2, and $n_L$= $n_c$ =$n_s$=1.

The objective of Fig. 3 is to connect the Rytov model with practical device geometry as applied in metamaterials presently. Accordingly, Fig. 3(a) shows an example grating membrane structure, enclosed by air, and its reflection spectrum mapped in wavelength versus grating thickness ($d_g$). The corresponding half-space grating structure used in the Rytov model is presented in Fig. 3(b). The effective refractive indices $n_m^{EMT}$ obtained by solving the exact Rytov Eq. (3) are shown in Fig. 3(c). The values of $n_m^{EMT}$ denote vertical components of the refractive indices $n_m$ that quasi-guided evanescent-wave diffraction orders see in the direction of propagation in the periodic medium. These evanescent diffraction orders excite lateral leaky Bloch modes that generate the guided-mode resonance. Comparing the rigorously-computed resonance map in Fig. 3(a) to Fig. 3(c) shows that no resonance occurs in the region where $n_1^{EMT}$=0. Moreover, using Eq. (6) with values of $n_m^{EMT}$ obtained by the exact Rytov formula, one can find the corresponding pairs of $n_m$ and $N_m$ satisfying the eigenvalue equation of the equivalent homogeneous slab waveguide [9]. Figure 3(d) depicts these values as a function of wavelength for an equivalent waveguide having a thickness of $d_g$= 0.4 μm. It can be inferred from this figure that the cutoff wavelength occurs when the refractive index of the waveguide reaches $n_m$=$n_{air}$=1 at which point the waveguide vanishes. Thus, at the cutoff wavelength, the refractive index contrast becomes zero such that no waveguide mode can be supported. For the grating design of Fig. 3(a) with $d_g$= 0.4 μm, the resonance manifests as a reflection peak at $\lambda_{Res}$= 1.251 μm as shown in Fig. 3(e). At the resonance wavelength, one can compute, with rigorous coupled-wave analysis [34], the amplitudes of the coupled diffracted orders and simulate the magnetic-field distribution as shown in Figs. 3(f) and 3(g), respectively. It is clearly illustrated that the dominant contribution to the internal modal field, whose cross-section is shown in Fig. 3(f), is due to the evanescent diffraction orders with amplitudes $S_{\pm 1}$ which is also completely consistent with the total magnetic-field distribution illustrated in Fig. 3(g). Interestingly, this point can be predicted and explained directly via Fig. 3(c); as the resonance wavelength fall below the cutoff wavelength of the first diffracted order ($\lambda_c^1$), we would expect the first diffracted orders $S_{\pm 1}(z)$ to be responsible for the resonance because it is this order that experiences $n_1^{EMT}$.

## 3. Rytov solutions for cutoff wavelengths

Knowing the values for the cutoff wavelengths is important to distinguish the deep-subwavelength and resonant-subwavelength regions. Moreover, the cutoff wavelengths define the spectral location where a new evanescent diffraction order, with attendant lateral Bloch-mode excitation, enters and begins to participate in the resonance dynamics. The cutoff wavelengths occur when the vertical effective refractive index of a propagative order vanishes (i.e., $n_m^{EMT} = 0$). The semianalytical Rytov formulas can be used to determine the first and higher cutoff wavelengths for any one-dimensional lattice. Therefore, plugging $n_{TE}^{EMT} = 0$ into the exact Rytov formulation, for example Eq. (2) for TE polarization, yields

$$n_L \tan\left[\frac{\pi\Lambda}{\lambda}(1-F)n_L\right] = -n_H \tan\left[\frac{\pi\Lambda}{\lambda}F n_H\right] \tag{7}$$

In general, there is no analytical solution for this equation. However, here we show that for specific design parameters one can straightforwardly and analytically calculate the cutoff wavelengths for each diffracted order. This works when the arguments of the tangent functions on each side of Eq. (7) become identical

$$(1-F)n_L = F n_H \tag{8}$$

Once this condition is satisfied, Eq. (7) holds for the values of the tangent arguments equal to $m\pi/2$ (m=1, 2, 3, …) which results in closed-form, simple analytical solutions

$$\lambda_c^m = \frac{2}{m}\Lambda F n_H \tag{9}$$

giving the cutoff wavelength for each diffraction order. All photonic lattices supporting guided-mode resonance admit at least the first evanescent diffraction order. Thus, with m=1, we get $\lambda_c^1 = 2\Lambda F n_H$. This is a remarkable canonical result. From Eq. (8), appropriate fill factors satisfying these solutions are F=$n_L$/($n_L$+$n_H$). These values of F are therefore reasonable for experimental realization. In the subwavelength regime, to ensure that at least one resonance arises from the $m^{th}$ diffraction order, the Rayleigh wavelength should be smaller than the cutoff wavelength (i.e., $\lambda_R < \lambda_c^m$). This yields a constraint F > $n_s$/2$n_H$ for m=1. Previously, Lalanne et al. obtained a numerical solution for $\lambda_c^1$ and pointed out its analogy with the Rayleigh wavelength [10].

In this spirit, one can engineer the spectral response and the number of diffracted orders at work by appropriately choosing the values of $\lambda_R$ and $\lambda_c^m$ for grating design. For example, the grating design depicted in Fig. 4(a) having parameters F=1/3, Λ=1 µm, $n_H$=2, and $n_L$=1 satisfies Eq. (8). Thus, the cutoff wavelengths for each evanescent diffracted order can be obtained analytically as expressed in Eq. (9) which returns values of $\lambda_c^1$=4/3~1.33 µm and $\lambda_c^2$=2/3~0.66 µm for the first two orders. The Rayleigh wavelength of this design is $\lambda_R$=1 µm which is smaller than the first order cutoff wavelength $\lambda_c^1$=4/3~1.33 µm.

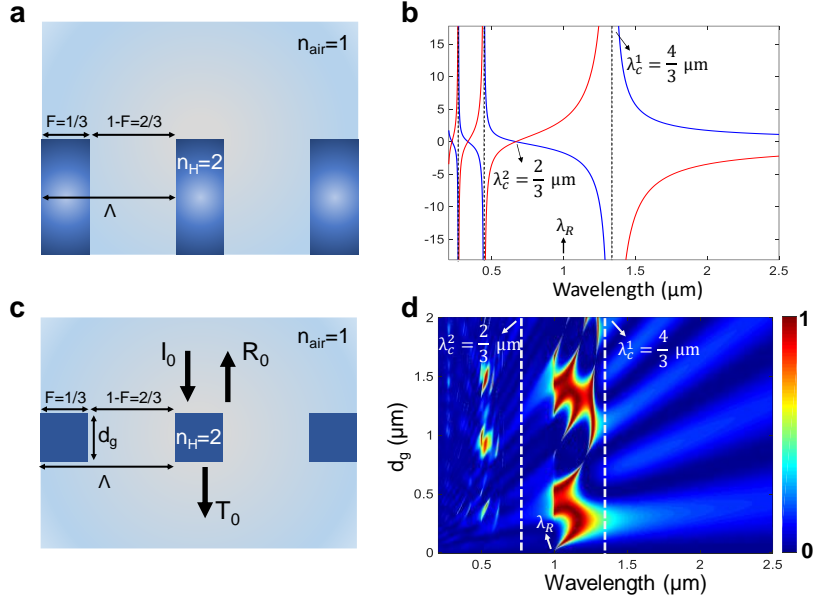

Fig. 4. Grating design with parameters satisfying conditions for analytic analysis. (a) Schematic of the half-space grating model, (b) graphical solution of Eq. (7) to find the cutoff wavelengths of the propagative diffractive order in the grating region, (c) schematic of a grating membrane with finite thickness, and (d) corresponding reflection map as a function of grating thickness ($d_g$).

To validate the accuracy of our method, it is seen in Fig. 4(b) that graphical solutions of Eq. (7), give the exact same values as obtained analytically by Eq. (9). Figure 4(c) shows a schematic of a grating membrane surrounded by air ($n_{air}$=1) with finite thickness of $d_g$. For this design, the cutoff wavelengths shown by dashed lines in Fig. 4(d), which is a RCWA-simulated reflection map, are in full agreement with the analytical cutoff values.

### 4. Relevance of Rytov's formulation to resonance device design

In this section, we show that the Rytov effective refractive indices are directly applicable to design of periodic photonic devices, including metamaterials and metasurfaces. Their deployment fully supports prior explanations of resonance device physics in terms of lateral leaky Bloch modes and guided-mode resonance [30-32]. Their existence and spectral expressions are not consistent with resonance effects caused by local modes including Fabry-Perot resonance or Mie scattering [35, 36]. Here, we treat example devices whose spectra and functionality are directly explainable using the Rytov indices.

*4.1 Wideband resonant reflector*

One particularly useful device is the wideband resonant reflector in which nanopatterned design provides high reflectivity approaching 100% over a wide wavelength range [31, 37]. Numerous studies have addressed these compact, often single-layer, reflectors both theoretically and experimentally for various optical wavebands [10, 32, 36-39]. Here, we apply the Rytov indices to substantiate the physical basis for the wideband reflection behavior. In this context, the half-space grating structure with parameters shown in Fig. 5(a) is considered for the analysis. Corresponding roots of the exact Rytov equations for both TM and TE polarization states are found and the results are shown in Figs. 5(b) and 5(c), respectively. Similar curves were obtained by Lalanne et al. [10] using an RCWA-based numerical algorithm. For the TM case shown in Fig. 5(b), there are two significant points to be considered. The first one concerns the values of the cutoff wavelengths for each guided diffracted order and the second pertains to the shape of the index curves. For instance, it is illustrated in Fig. 5(b) that in the wavelength range of 1.25 μm to 3 μm, beyond the cutoff wavelength of the second order, only $n_0^{EMT}$ and $n_1^{EMT}$ exist in the effective refractive index diagram. Consequently, these two orders with m=0 and m=1 are responsible for all important spectral properties. Furthermore, it is seen that the slopes of the curves are almost identical with both curves varying monotonically in a wide wavelength range depicted by the gray region in Fig. 5(b). This is a key point to achieve wideband reflector response as the wavelength dependent phase difference (Δφ) accumulated in the z direction between these two orders at work is defined by $\Delta\varphi = (2\pi/\lambda_0)(n_0^{EMT}(\lambda_0) - n_1^{EMT}(\lambda_0))d_g$ which is proportional to the effective refractive-index difference of the first two orders obtained by the exact Rytov expression. Therefore, our method enables us to predict whether to expect a wideband reflector behavior from a one-dimensional grating structure, simply by calculating effective refractive index graphs without performing any rigorous numerical simulations. The closed-form Rytov formulas might thus substantiate efficient design methods. Applying this approach to Fig. 5(c), it is seen directly that no wideband reflection response will arise out of this design for TE polarization as the slopes of the two curves differ significantly. To confirm our hypothesis, we performed RCWA-based simulations for the structure shown in Fig. 5(d). Simulated zero-order reflection maps of this grating design for TM and TE cases are shown in Figs. 5(e) and 5(f), respectively. These maps validate our predictions of wideband reflection response occurring in TM polarization but no wideband reflection response for the TE case. Wideband reflectors are related to the regions with dark red colors in a wide wavelength range. These appear in the TM map at some specific grating thicknesses $d_g$ which provide an appropriate phase difference (completely in phase) for high reflection since $\Delta\varphi \propto d_g$. Explanation of wideband resonance reflection applying the spectral phase pertinent to similar, albeit numerically-simulated, effective indices was first provided by Lalanne et al. [10].

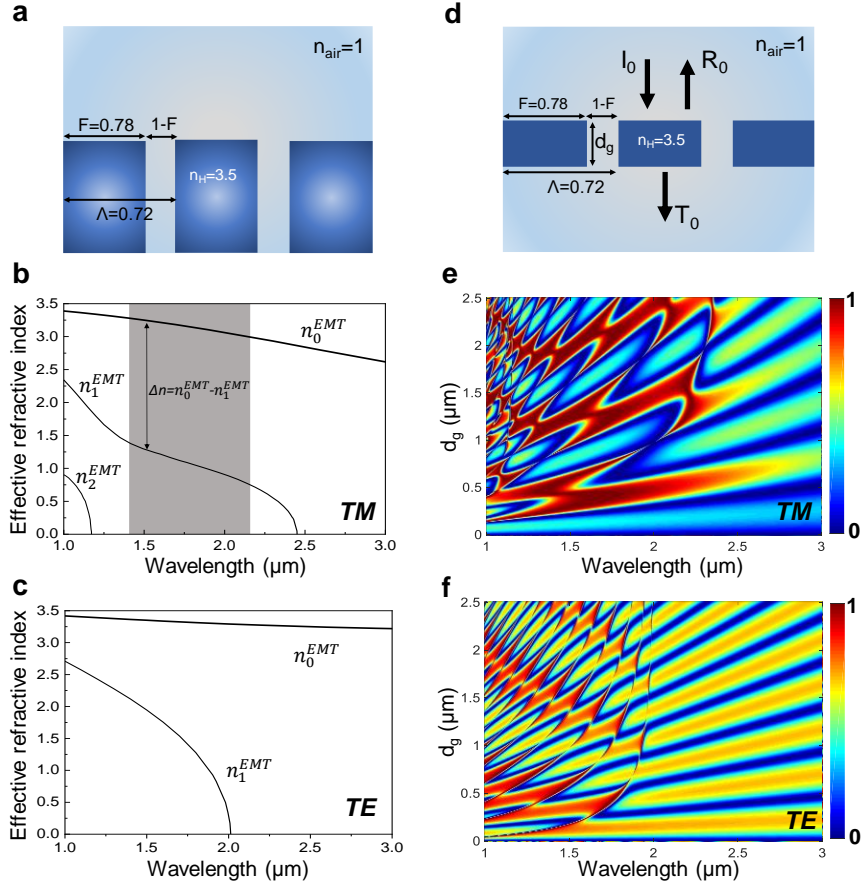

Fig. 5. An example demonstrating the use of the Rytov indices for design of a wideband resonant reflector. (a) Schematic of the half-space model. Calculated effective refractive indices using the Rytov formalism for (b) TM-polarization, and (c) TE-polarization states. (d) A schematic of a corresponding grating membrane with parameters $\Lambda= 0.72$ μm, F=0.78, $n_H$=3.5, and $n_L = n_{air}$=1. Simulated reflection maps in wavelength versus grating thickness ($d_g$) pertinent to normally-incident (e) TM-polarized, and (f) TE-polarized light. In the maps, dark red color implies $R_0$ approaching 1.

*4.2 Guided-mode resonant bandpass filter*

Another important grating-based optical device is the sparse, single-layer bandpass filter (BPF) exhibiting low transmission sidebands and high-efficiency narrow-band transmission peak [40-43]. Low transmission sidebands and a transmission resonance peak correspond to a wideband high-reflection background and a reflection resonance dip, respectively. To study this device type, a half-space model and the corresponding calculated Rytov refractive indices are shown in Fig. 6(a). We choose the grating parameters to satisfy Eq. (8) to analytically obtain the cutoff wavelengths. As in the explanation of the wideband reflector, similarity in the slopes of the $n^{EMT}$ curves enables an appropriate phase difference to obtain high reflectivity at a specific device thickness.

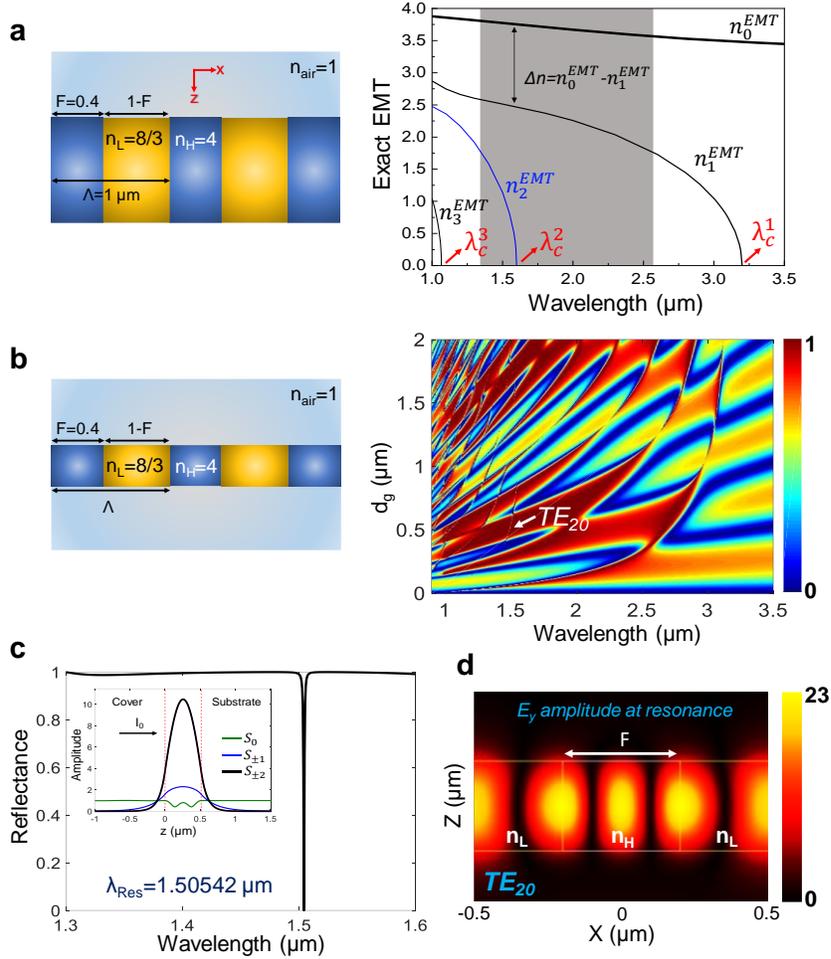

Fig. 6. Rytov indices in bandpass filter analysis and design. The example filter works in TE polarization with parameters of $\Lambda= 1$ μm, F=0.4, $n_H$=4, and $n_L$=8/3, and $n_{air}$=1. (a) Schematic of the Rytov half-space model and calculated effective refractive indices. (b) Schematic of the attendant grating membrane with finite thickness and its simulated reflection map as a function of grating thickness ($d_g$). (c) Bandpass filter response of the device with grating thickness of $d_g$=0.51 μm. Inset in (c) shows the amplitudes of the coupling diffracted orders at the resonance wavelength of $\lambda_{Res}$=1.50542 μm. (d) Electric field distribution at resonance exhibiting a $TE_{20}$ profile.

Thus, we expect a wideband reflector response in the wavelength range where $\Delta n/\lambda_0$ is relatively constant. This condition prevails in the gray region of the EMT graph of Fig. 6(a). This figure is significantly different from Fig. 5(b) in that the gray region in Fig. 5(b) contains only $n_0^{EMT}$ and $n_1^{EMT}$ whereas the gray region in Fig. 6(a) encompasses $n_0^{EMT}$, $n_1^{EMT}$ and $n_2^{EMT}$. As $n_0^{EMT}$ and $n_1^{EMT}$ are responsible for a wideband reflection background, bringing the second order $n_2^{EMT}$ to work will manifest as a reflection dip resonance feature in the optical spectrum because it exists within a region of total reflection. Figure 6(b) shows a schematic of the membrane version of the half-space grating design of Fig. 6(a) and its reflection map as a function of grating thickness. The reflection map agrees well with the analytic solutions for the cutoff wavelengths and with the number of orders at work experiencing $n_0^{EMT}$, $n_1^{EMT}$, and $n_2^{EMT}$. The resonance feature predicted based on the existence of the $n_2^{EMT}$ curve in the effective refractive index graph is marked as $TE_{20}$ in the reflection map of Fig. 6(b). Figure 6(c) confirms a bandpass filter response having a wideband high reflection background. At the reflection dip wavelength, the inset in Fig. 6(c) reveals that the second evanescent diffraction order m=2 is dominant showing that a non-zero $n_2^{EMT}$ is key to realizing a bandpass filter. The electric field distribution at the resonance wavelength shown in Fig. 6(d) furthermore indicates $TE_{20}$ response (fundamental mode excited by the second evanescent order) consistent with our model. In summary, the Rytov treatment of the resonant BPF is fully consistent with, and supports, prior descriptions of BPF physics [40, 41, 43].

### 4.3 Guided-mode resonance polarizer

The linear resonant polarizer is the final device example presented. In the past, it has been shown that ultra-compact polarizers with high extinction ratios are realizable with resonant gratings [31, 44, 45]. Treating here a known polarizer [45], the design schematic is shown in Fig. 7(a) displaying a small fill factor (F=0.1) with $n_H$=3.5 embedded in a medium with refractive index of 1.5 in a way that $n_L$= $n_C$= $n_S$=1.5 under normal incidence. Figure 7(b) shows computed λ-$d_g$ reflection maps for TE and TM polarization states. At the specific thickness of the grating denoted by the dashed line, TE polarization exhibits high reflection while TM reflectance is suppressed. Reflectance spectra for a grating with thickness $d_g$=0.54 μm as shown in Fig. 7(c) reveal a good polarizing response in a wavelength range of 1.3-1.5 μm. To elucidate the polarization behavior in the Rytov picture, we calculate TE and TM Rytov indices as presented in Fig. 7(d). Again, the parallelism of the $n_0^{EMT}$ and $n_1^{EMT}$ curves in TE polarization enables a wideband reflector response. In contrast, for TM polarization in Fig. 7(d) in the working range of the polarizer, only $n_0^{EMT}$ exists. Consequently, we see that no guided-mode resonance features will occur in the TM case consistent with the simulated reflection map in Fig. 7(b).

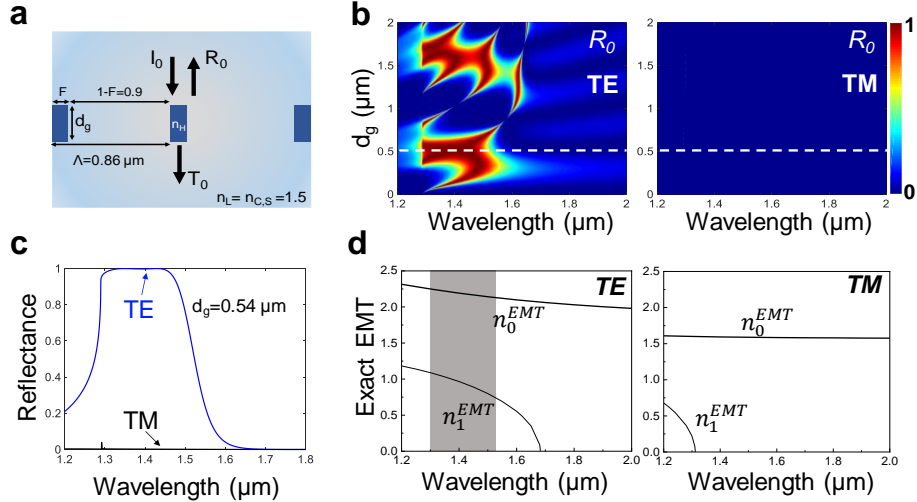

Fig. 7. Rytov analysis of a sparse grating polarizer with parameters Λ= 0.86 μm, F=0.1, $n_H$=3.5, and $n_L$=1.5 extracted from Ref. [45]. (a) Schematic of the design with finite grating thickness of $d_g$, (b) RCWA-based λ-$d_g$ reflection map for TE and TM polarization states, (c) reflectance spectra with $d_g$=0.54 μm, and (d) exact Rytov effective refractive index diagram for TE and TM cases.

## 5. Conclusion

In summary, we present Rytov refractive indices obtained by solving the exact Rytov formulation initially derived in 1956. In contrast to most past work where effective medium theory returns a single effective index for a given device, the full formalism provides multiple solutions based on the multiple roots inherent therein. Thus EMT for a one-dimensional periodic grating results in multiple effective refractive indices that are wavelength dependent.

The full set of Rytov indices is directly applicable to design of periodic photonic devices, including metamaterials and metasurfaces. Their manifestation fully supports explanations of resonance device physics in terms of lateral leaky Bloch modes and guided-mode resonance. The cutoff wavelengths of the evanescent diffraction orders define their spectral region of dominance and interaction. The spectral slope of the Rytov indices predicts spectral ranges across which the reradiated Bloch modes will be in phase or out of phase. Thus, for example, it is possible to predict whether to expect a wideband reflector behavior from a one-dimensional grating structure simply by calculating effective refractive index graphs without performing any rigorous numerical simulations. The closed-form Rytov formulas might thus substantiate efficient design methods. The fact that the cutoff wavelengths are directly embedded in the formulation enables definition of the dividing line between the resonance subwavelength region and the deep-subwavelength region based on the cutoff wavelength of the first evanescent diffraction order. This important transition point is always numerically available via the Rytov formulation. In a special case, we find that the transition wavelength is given by $\lambda_c^1 = 2Fn_H\Lambda$ which is directly comparable to the Rayleigh wavelength $\lambda_R = n_S\Lambda$ that defines transition from the non-subwavelength to the subwavelength regime. We successfully apply the Rytov formalism to reliably describe the behavior of various optical devices, such as wideband reflectors, resonant bandpass filters, and guided-mode resonance polarizers. Rigorous numerical results support all of our explanations and predictions. Future studies might investigate and extend the methods of this study to more complex lattices such as those with multipart unit cells and two-dimensional periodic metasurfaces. The utility and precision with which the simple Rytov formalism applies to resonant metamaterials is an important discovery that will come as a surprise to most and will count as a major advance in the development of the field.

**Funding.** This research was supported, in part, by the UT System Texas Nanoelectronics Research Superiority Award funded by the State of Texas Emerging Technology Fund as well as by the Texas Instruments Distinguished University Chair in Nanoelectronics endowment. Additional support was provided by the National Science Foundation (NSF) under Awards No. ECCS-1606898, ECCS-1809143, and IIP-1826966.

**Disclosures.** The authors declare no conflicts of interest.